\documentclass[journal,10pt,twocolumn]{IEEEtran}
\usepackage{graphicx,cite}
\usepackage{microtype}
\usepackage{amsmath}
\usepackage{amsfonts}
\usepackage{amssymb}
\usepackage{xcolor}
\newcommand{\rev}[1]{#1}
\usepackage{makecell}
\usepackage{multirow}
\usepackage{bm}
\usepackage{colortbl}
\usepackage{booktabs}
\usepackage{tabularx}
\usepackage{url}
\usepackage[hidelinks]{hyperref}
\usepackage{tikz}
\usetikzlibrary{positioning,arrows.meta,calc}
\graphicspath{{Images/}}
\tikzset{
 blk/.style={draw, rounded corners=2pt, fill=blue!6, align=center, font=\footnotesize, inner sep=2.5pt, minimum height=7mm},
 dat/.style={draw, rounded corners=2pt, fill=gray!8, align=center, font=\footnotesize, inner sep=2.5pt, minimum height=7mm},
 sig/.style={-{Latex[length=1.6mm]}, semithick},
 lbl/.style={font=\scriptsize, align=center},
 sumn/.style={draw, circle, inner sep=0pt, minimum size=4mm, font=\footnotesize},
}
\newcommand{\FigPGML}{%
\begin{tikzpicture}[x=1mm,y=1mm]
\node[dat] (x)    at (4,0)    {$\mathbf v_{dq}[k]$\\$\mathbf i_{dq}[k\!-\!1]$\\$\omega_r[k\!-\!1]$};
\node[blk] (inc)  at (30,0)   {Increment\\network $\mathcal{N}_{\theta}$\\(5--28--28--3)};
\node[sumn] (sum) at (59,0)   {$+$};
\node[blk] (spd)  at (30,17)  {Speed unit: trapezoidal\\swing equation \eqref{eq:linear}};
\draw[sig] (x) -- (inc);
\draw[sig] (inc) -- node[lbl,above]{$\Delta\hat{\mathbf i}_{dq}[k]$} (sum);
\draw[sig] (sum) -- ++(14,0) node[lbl,above,pos=0.6]{$\mathbf i_{dq}[k]$};
\draw[sig] (59,-9) node[lbl,below]{$\mathbf i_{dq}[k\!-\!1]$} -- (sum);
\draw[sig] (inc.north east) -- ++(6,0) |- node[lbl,right,pos=0.25]{$\hat T_e[k]$} (spd.east);
\draw[sig] ($(spd.north)+(-12,7)$) node[lbl,left]{$T_m$} -| ($(spd.north)+(-12,0)$);
\draw[sig] (spd.west) -- ++(-9,0) node[lbl,above,pos=1]{$\omega_r[k],\,\theta_e[k]$};
\end{tikzpicture}}

\newcommand{\FigStages}{%
\resizebox{\columnwidth}{!}{\begin{tikzpicture}[x=1mm,y=1mm]
\tikzset{st/.style={font=\footnotesize\itshape, anchor=west}}
\foreach \y/\h/\s in {0/0/A, -21/0/B, -42/4/C, -66/4/D, -91/4/E}{
  \node[st] at (-3.5,\y+8) {Stage \s};
  \draw[gray!60, rounded corners=2pt] (-4,\y+11) rectangle (88,\y-8-\h);
}
\node[dat] (rA) at (71,0) {Referencer: PMSM1\\linear model \eqref{eq:linear}};
\node[blk] (iA) at (40,0) {IIL\\training};
\node[blk] (mA) at (13,0) {Increment\\network $\mathcal{N}_{\theta}$};
\node[dat] (rB) at (71,-21) {Referencer: PMSM2\\damper model \eqref{eq:damper}};
\node[blk] (iB) at (40,-21) {IIL\\training};
\node[blk] (mB) at (13,-21) {PMSM3\\(7--30--30--5)};
\node[dat] (rC) at (71,-45) {Referencer: PMSM4\\shifted parameters};
\node[blk] (iC) at (40,-45) {Retrain $\mathcal{N}_{\theta}$;\\identify $\bm\theta_p$};
\node[blk] (mC) at (13,-41) {PMSM6};
\node[blk, fill=gray!10] (bC) at (13,-50) {PMSM5};
\node[dat] (rD) at (71,-69) {Referencer: PMSM7\\FEA map, look-up};
\node[blk] (iD) at (40,-69) {IIL training\\of $\mathcal{N}_{\theta}$ and $\mathcal{P}_{\psi}$};
\node[blk] (mD) at (13,-65) {PMSM9};
\node[blk, fill=gray!10] (bD) at (13,-74) {PMSM8};
\node[dat] (rE) at (71,-94) {Referencer: M1 bench\\M2: controlled load};
\node[blk] (iE) at (40,-94) {Real-time IIL\\measured data};
\node[blk] (mE) at (13,-94) {PMSM11\\Residual PGML\\($100~\mu$s)};
\foreach \s in {A,B,C,D,E}{\draw[sig] (r\s) -- (i\s);}
\draw[sig] (iE) -- (mE);
\draw[sig] (iA) -- (mA); \draw[sig] (iB) -- (mB);
\draw[sig] (iC.west) -- ++(-5,0) |- (mC.east); \draw[sig] (iC.west) -- ++(-5,0) |- (bC.east);
\draw[sig] (iD.west) -- ++(-5,0) |- (mD.east); \draw[sig] (iD.west) -- ++(-5,0) |- (bD.east);
\draw[sig, dashed] (mA.west) -- (-7,0) |- (mC.west);
\draw[sig, dashed] (-7,-41) |- (mD.west);
\draw[sig, dash dot, blue!65!black] (mD.south) -- ++(0,-1) -- (-10,-69.5) |- (mE.west);
\end{tikzpicture}}}

\newcommand{\MPAEcell}[2]{\cellcolor{blue!#1}\textcolor{black}{\strut #2}}

\begin{document}
\title{Physics-Guided Inferencer-in-the-Loop\\Digital-Twin for Real-Time Adaptive\\Emulation of Nonlinear PMSMs}

\author{Songyang~Zhang,~\IEEEmembership{Member,~IEEE,} Yuzhong~Zhang,~\IEEEmembership{Member,~IEEE}, and Venkata~Dinavahi,~\IEEEmembership{Fellow,~IEEE}
	\thanks{This work is supported by the Natural Sciences and Engineering
		Research Council of Canada (NSERC), Mitacs Accelerate program, and RTDS Technologies Inc. (Corresponding author: Songyang Zhang)}
	\thanks{S. Zhang was with the Department of Electrical and Computer Engineering, University of Alberta, Edmonton, AB T6G 2V4, Canada. He is now with RTDS Technologies Inc., Winnipeg, MB R3T 2E1, Canada (e-mail: zhang.songyang@ualberta.ca).}
	\thanks{Y. Zhang and V. Dinavahi are with the Department of Electrical and Computer Engineering, University of Alberta, Edmonton, AB T6G 2V4, Canada (e-mail: yuzhong@ualberta.ca; dinavahi@ualberta.ca).}}
\maketitle

\begin{abstract}
A real-time digital-twin (RTDT) of a permanent magnet synchronous machine (PMSM) allows drive controllers to be tested in hardware-in-the-loop (HIL) before they meet the real machine. \rev{Its machine model rests on identified parameters or flux look-up tables (LUTs) from finite-element analysis (FEA) or characterization tests; the maps are regenerated offline when the machine changes, and online estimators track a few parameters such as resistance and PM flux.} This article makes the executed model learnable from reference data. \rev{A physics-guided machine learning (PGML) network advances the currents and predicts torque in one step, trained with analytical one-step labels. In the proposed Inferencer-in-the-Loop (IIL) architecture, a GPU retrains the network while a field-programmable gate array (FPGA) executes the last accepted weights.} For damper circuits, shifted parameters, and an FEA-calibrated saturated machine, one kernel form reaches average normalized mean absolute errors (NMAEs) of 0.19\%--0.67\% at measured FPGA \rev{kernel latencies of \rev{0.30--0.34}~$\mu$s, and retraining for a shifted machine nears its final loss within 94~s. On a \rev{magnetically linear 200-W drive}, a residual form corrects the analytical step and learns its physical parameters from operating data; each step \rev{takes} under 2.5~$\mu$s, 20 times faster than real time \rev{and ahead of the physical drive's measurements} even at 20-kHz sampling. On six held-out runs (1193~s), its $i_q$ error is 5.9\%--9.4\% of measured RMS, against 10.4\%--13.2\% for an independently fitted grey-box model, and over 30~min of loaded heating its online-updated heads track a 4.1\% winding-resistance rise with an RMS error of 0.44~m$\Omega$.}
\end{abstract}


\section{Introduction}
Permanent magnet synchronous machines (PMSMs) are central to electric-vehicle, ship-propulsion, and industrial drives. Hardware-in-the-loop (HIL) testing uses real-time digital-twins (RTDTs) to reproduce their currents, torque, and speed under saturation, cross-saturation, demagnetization, and faults \cite{007Espinosa2010,007Ursic2019,007Zhang2018}, \rev{so that control and protection, including fault handling, can be verified without risk to the hardware.} An RTDT is only as useful as its model is close to the installed machine, whose parameters vary between units and drift with temperature.

Electromagnetic transient simulators represent PMSM nonlinearity through piecewise models and current-dependent look-up tables (LUTs) derived from finite element analysis (FEA), including field-programmable gate array (FPGA) implementations \cite{Dufour2007,Chen2015,Fan2025,Li2026}. Extensions capture iron losses and thermal effects \cite{Luo2016,Scheer2022}, multi-phase coupling \cite{Zou2024}, bivariate flux relations \cite{Sun2020}, and cross-magnetization \cite{Dehkordi2026}. Drive-based identification gives the measured counterpart, estimating magnetic models in dedicated tests \cite{Armando2013,Pellegrino2015,Hinkkanen2017}, resistance and PM flux \cite{Odhano2019,Liu2013}, and temperature \cite{Kirchgassner2021,Kirchgassner2023TNN}. These models are accurate and interpretable. \rev{When the machine changes, its maps are recomputed from new FEA runs or tests, online identification adjusts resistance and PM flux in a fixed model structure, and characterizing the full current plane requires high-current, high-temperature points that are costly to reach;} effects outside that structure, such as inverter nonlinearity, need separate models.

Machine learning (ML) represents magnetic characteristics compactly \cite{007Li2023,007Serrano2023}, has entered real-time simulation \cite{RTACDCbook,RTHEML}, and incorporates governing equations through physics-informed training \cite{RAISSI2019,Zhang2025SMR}. For electric machines, networks represent FEA flux tables \cite{Bai2020}, learn flux-linkage models from drive data \cite{Sevcik2023,Sevcik2025}, and preserve magnetic reciprocity by construction \cite{LiGradNet2026}. These networks are trained offline as components of an equation-based model. A network that is itself the executed real-time step, retrained from simulated or measured data while the emulator runs, has received less attention.

\rev{The goal here is an emulator whose executed step can be relearned from whatever reference is at hand, be it an equation-based model, an FEA-calibrated model, or signals recorded on a drive, while the FPGA keeps its fixed step. In the proposed Inferencer-in-the-Loop (IIL) workflow, the FPGA executes a compact physics-guided machine learning (PGML) network that maps the present state and voltages to the next-step current increment and torque. One network structure covers damper dynamics and FEA-calibrated saturation, and the analytical one-step model serves as a training label rather than as the executed model. Retraining runs on a GPU, and the FPGA switches to new weights only after they pass validation. On a measured drive, where the equations are only approximate, the network instead corrects an analytical step whose physical parameters are learned with it. Because the executed step itself is learned, one update path covers parameter drift, unmodeled inverter effects, and saturation.} The contributions are as follows:

1) A current-increment PGML formulation whose executed step keeps one form and a cost of \rev{30--34} cycles from damper circuits to FEA saturation, and a residual extension that learns measured-drive dynamics and physical parameters from operating data without flux-map characterization.

2) An IIL architecture that separates GPU training, reference-data exchange, and deterministic FPGA execution, with validation governing model release, so an update leaves the controller interfaces and the real-time schedule unchanged. It extends our earlier IIL work \cite{IIL} to adaptive machine emulation.

3) Validation in four simulation stages with FPGA-measured latencies and on a 200-W drive, where the RTDT is 20 times faster than real time \rev{even at 20-kHz sampling}, outperforms a grey-box model on all test runs, and tracks real heating online.

Sections~\ref{sec:models}--\ref{sec:iil} present the machine models, the PGML formulations, and the IIL architecture; Sections~\ref{sec:results} and~\ref{sec:bench} give simulation and experimental validation; Section~\ref{sec:limits} discusses implications, and Section~\ref{sec:conclusion} concludes.

\section{PMSM Models and Model Variants}\label{sec:models}
\subsection{Per-Unit \texorpdfstring{$dq$}{dq}-Axis Models}
All simulation models are in per-unit (p.u.) form on the bases of Table~\ref{tab:pmsm_parameters}, with $\omega_B=2\pi\cdot60$~rad/s. With the stator flux linkages as states, the linear model can be described as:
\begin{equation}\label{eq:linear}
\left\{
\begin{aligned}
\frac{d\lambda_d}{dt} &= \omega_B\left(v_d - R_s i_d + \omega_r \lambda_q\right)\\
\frac{d\lambda_q}{dt} &= \omega_B\left(v_q - R_s i_q - \omega_r \lambda_d\right)\\
\lambda_d &= L_d i_d + \lambda_{pm}, \qquad \lambda_q = L_q i_q\\
T_e &= \lambda_d i_q - \lambda_q i_d\\
2H\frac{d\omega_r}{dt} &= T_e - T_m - B\omega_r, \qquad \frac{d\theta_e}{dt} = \omega_B\,\omega_r
\end{aligned}
\right.
\end{equation}
Here, $\omega_r$ is p.u. speed, $\theta_e$ is electrical angle, $T_e$ and $T_m$ are p.u. torques, and $B$ is the p.u. friction coefficient. The inertia constant $H=J_m\Omega_{mB}^2/(2S_B)$ is in seconds, with shaft inertia $J_m$ and base mechanical speed $\Omega_{mB}$. The electrical quantities use the usual $dq$ notation. Rotor damper circuits describe sub-transient dynamics,
\begin{equation}\label{eq:damper}
\left\{
\begin{aligned}
\frac{d\lambda_{kd}}{dt} &= -\omega_B r_{kd}\, i_{kd}, \qquad \frac{d\lambda_{kq}}{dt} = -\omega_B r_{kq}\, i_{kq}\\
\lambda_d &= L_d i_d + L_{md} i_{kd} + \lambda_{pm}\\
\lambda_{kd} &= L_{md} i_d + L_{kd} i_{kd} + \lambda_{pm}\\
\lambda_q &= L_q i_q + L_{mq} i_{kq}, \qquad \lambda_{kq} = L_{mq} i_q + L_{kq} i_{kq}
\end{aligned}
\right.
\end{equation}
with damper resistances $r_{kd}$, $r_{kq}$, self-inductances $L_{kd}$, $L_{kq}$, and $L_{md}=L_d-L_{ls}$, $L_{mq}=L_q-L_{ls}$; the currents follow by inverting \eqref{eq:damper}. In the simulation stages, all models use a fixed step $\Delta t=1~\mu$s, and the stator fluxes use a sequential first-order step:
\begin{equation}\label{eq:euler}
\begin{aligned}
\lambda_q[k] &= \lambda_q[k-1]+\Delta t\,\omega_B\big(v_q[k]-R_s i_q[k-1]\\
&\qquad\qquad\qquad-\omega_r[k-1]\lambda_d[k-1]\big),\\
\lambda_d[k] &= \lambda_d[k-1]+\Delta t\,\omega_B\big(v_d[k]-R_s i_d[k-1]\\
&\qquad\qquad\qquad+\omega_r[k-1]\lambda_q[k]\big).
\end{aligned}
\end{equation}
The $q$-axis update precedes the $d$-axis update, and the damper fluxes use forward Euler. Speed and angle use the trapezoidal rule. The analytical models advance flux states; the surrogate advances current states directly (Section~\ref{sec:pgml}).

\begin{table}[!t]
\centering
\scriptsize
\caption{Parameters of the Base Machine (PMSM1--PMSM6) and Drive}
\label{tab:pmsm_parameters}
\renewcommand{\arraystretch}{1.05}
\setlength{\tabcolsep}{2.5pt}
\begin{tabular}{lcc}
\toprule
\textbf{Parameter} & \textbf{p.u.} & \textbf{SI} \\ \midrule
Rated power $S_B$; line voltage $U_N$ (RMS) & -- & 60 kVA; 300 V \\
Base voltage $U_B$; current $I_B$ (peak) & 1 & 244.9 V; 163.3 A \\
Stator resistance $R_s$ & 0.0021 & 3.15 m$\Omega$ \\
Leakage inductance $L_{ls}$ & 0.064 & 0.255 mH \\
Inductances $L_d=L_q$ & 0.689 & 2.74 mH \\
PM flux linkage $\lambda_{pm}$ & 1.0 & 0.650 Wb \\
Inertia constant $H$; friction $B$ & --; 0.0005 & 0.25 s; -- \\ \midrule
Damper resistances $r_{kd}$; $r_{kq}$ & 0.055; 0.183 & 82.5; 274.5 m$\Omega$ \\
Damper inductances $L_{kd}$; $L_{kq}$ & 0.675; 0.725 & 2.69; 2.88 mH \\ \midrule
PMSM4: $L_d$; $L_q$; $\lambda_{pm}$ & 0.62; 0.56; 0.83 & -- \\
DC link $V_{dc}$; PWM carrier $f_c$ & -- & 540 V; 12 kHz \\
\bottomrule
\end{tabular}
\end{table}

\subsection{FEA-Based Nonlinear Parameter Maps}
\begin{figure*}[t]
	\centering
	\includegraphics[height=0.197\textwidth]{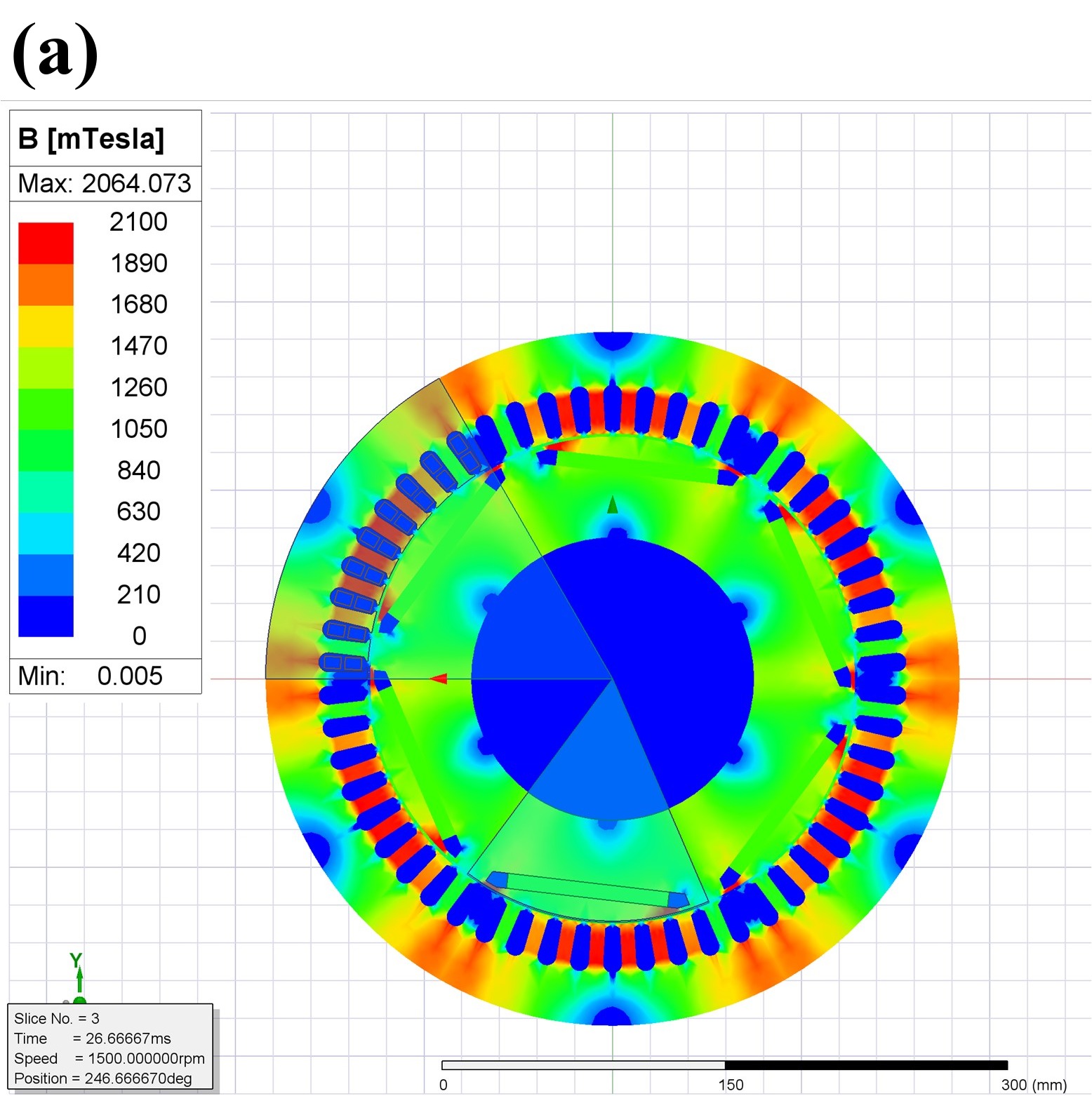}\hfill
	\includegraphics[height=0.197\textwidth]{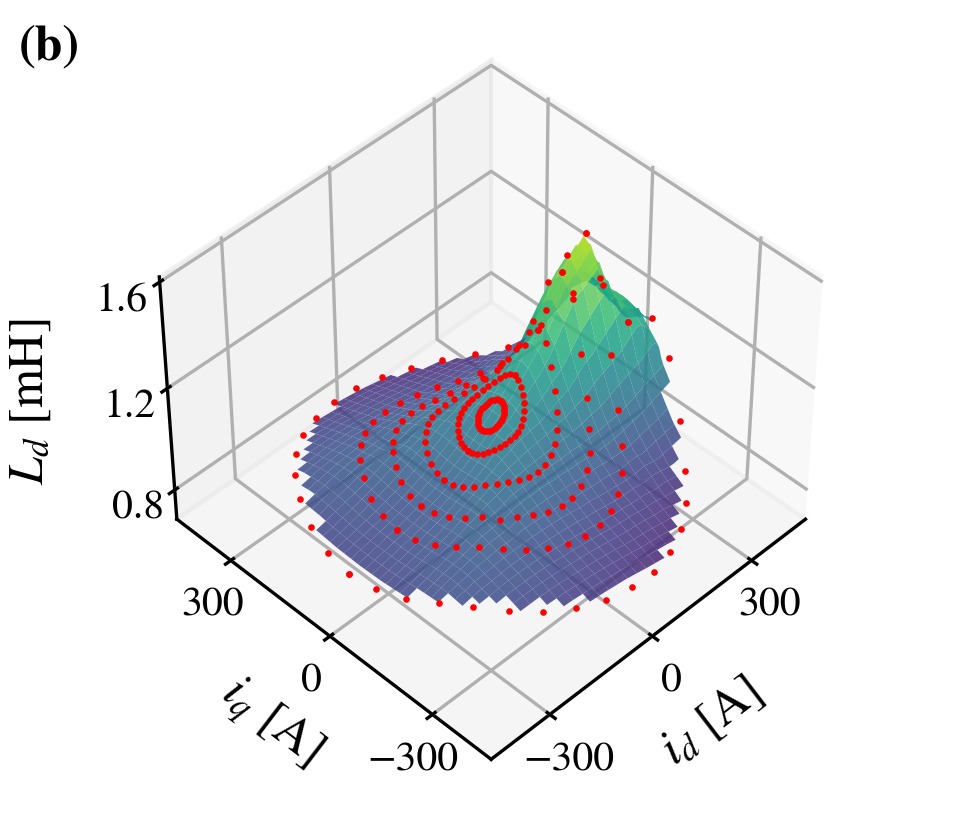}\hfill
	\includegraphics[height=0.197\textwidth]{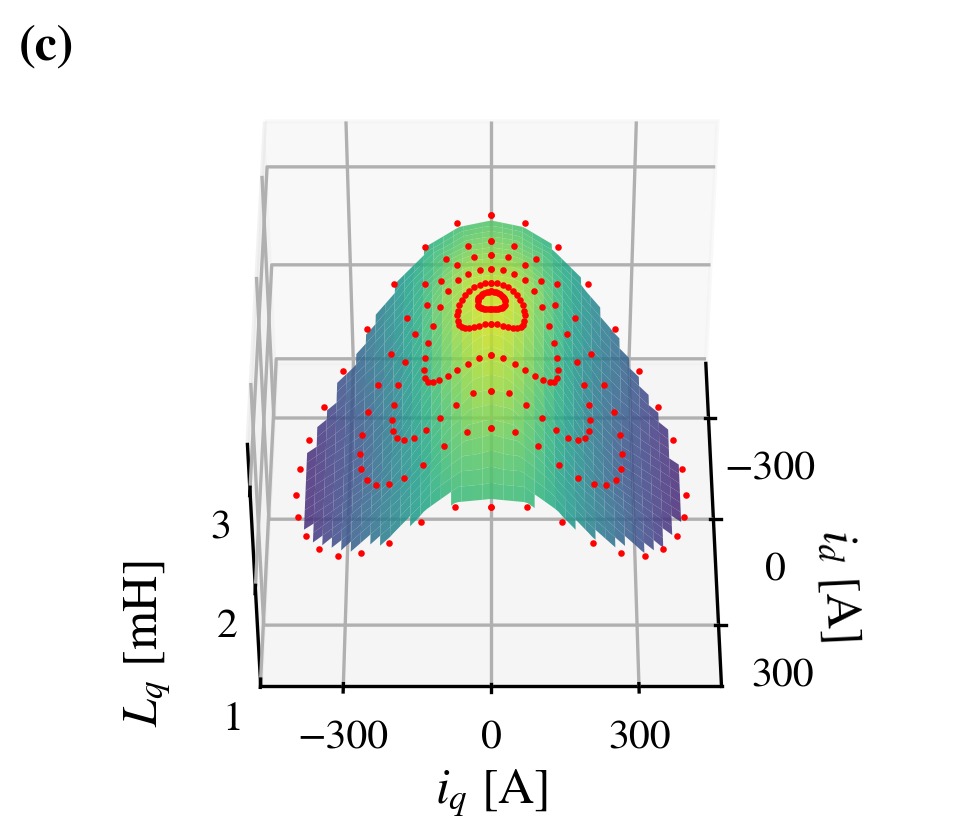}\hfill
	\includegraphics[height=0.197\textwidth]{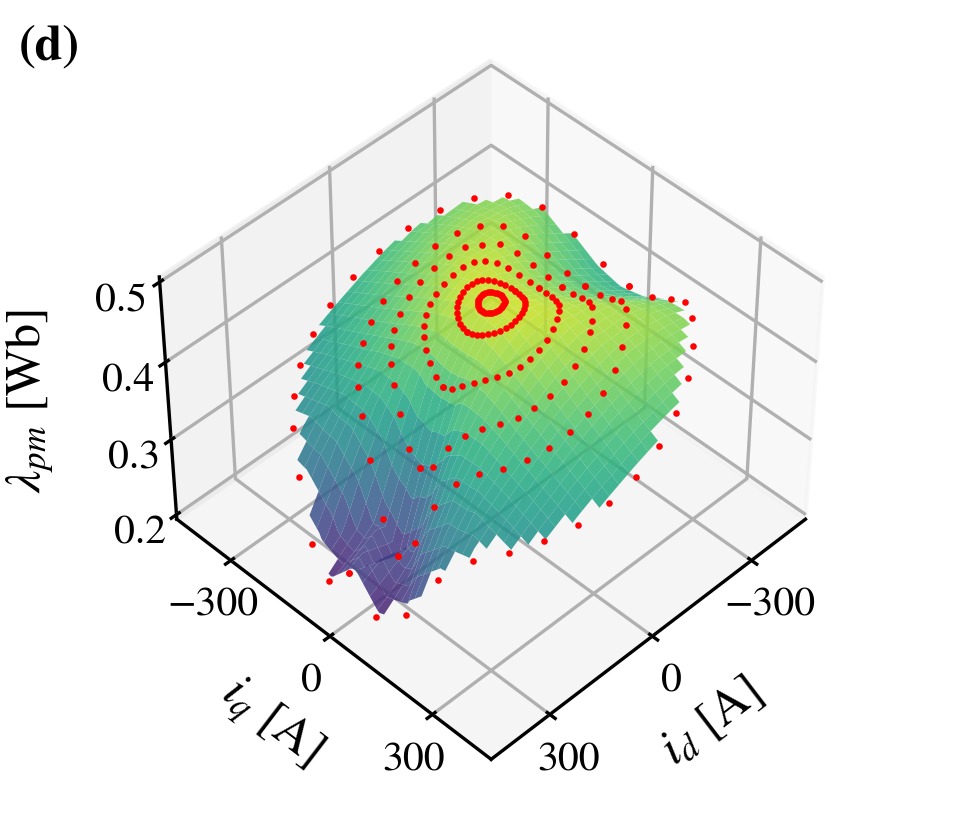}
	\caption{FEA-calibrated IPM machine: (a) Ansys Maxwell model with the flux density at 1500~rpm; apparent parameters (b)~$L_d(i_d,i_q)$, (c)~$L_q(i_d,i_q)$, and (d)~$\lambda_{pm}(i_d,i_q)$, where red dots mark the FEA points ($i_q\geq0$) and their mirror images. \label{fig:fea}}
\end{figure*}

The machine of the last stage is an interior permanent magnet (IPM) machine modeled in Ansys Maxwell (Fig.~\ref{fig:fea}(a)). Maxwell provides the $dq$ flux linkages and the PM flux linkage $\lambda_{pm}$ on a polar grid of six current amplitudes (0.2--3~p.u.) and 19 current angles ($0^\circ$--$180^\circ$, i.e., $i_q\geq0$), with $I_B=136$~A, $L_B=3.38$~mH, and $\lambda_B=0.459$~Wb. The apparent inductances follow as $L_d=(\lambda_d-\lambda_{pm})/i_d$ and $L_q=\lambda_q/i_q$, so that the flux--current relation of \eqref{eq:linear} holds exactly at every FEA point and the maps can replace its constants. Over this range, $L_d$ varies between 0.79 and 1.54~mH, $L_q$ between 1.22 and 3.23~mH, and $\lambda_{pm}$ between 0.22 and 0.48~Wb, reflecting strong saturation and cross-saturation (Fig.~\ref{fig:fea}(b)--(d)).

 The maps are normalized with their own p.u. base, re-gridded onto a uniform $13\times37$ grid in $(i_q,i_d)$, and evaluated at $|i_q|$; the FEA machine keeps the resistance, inertia, and controller of the base machine. Each FEA model step advances the fluxes by \eqref{eq:euler}, recovers the currents with the previous-step parameters, and then evaluates the maps, holding boundary values beyond the grid. PMSM7 uses nearest-neighbor look-up, as LUT-based emulators do, and PMSM10 bilinear interpolation; each stores $3\times13\times37=1443$ values. PMSM8 and PMSM9 learn from PMSM7 and are evaluated against PMSM10. Table~\ref{tab:pmsm_variants} summarizes the compared models.

\begin{table}[!t]
\centering
\footnotesize
\caption{Definition of PMSM Model Variants}
\label{tab:pmsm_variants}
\renewcommand{\arraystretch}{1.08}
\begin{tabularx}{\columnwidth}{lX}
\toprule
\textbf{Model} & \textbf{Description} \\ \midrule
PMSM1 & Linear model \eqref{eq:linear} of the base machine; stage-A referencer \\
PMSM2 & Base machine with dampers \eqref{eq:damper}; stage-B referencer \\
PMSM3 & PGML surrogate with dampers, trained against PMSM2 \\
PMSM4 & Base machine with reduced $L_d$, $L_q$, $\lambda_{pm}$; stage-C referencer \\
PMSM5 & Linear model with the parameters identified in stage C \\
PMSM6 & PGML surrogate retrained against PMSM4 (stage C) \\
PMSM7 & FEA machine, nearest-neighbor look-up; stage-D referencer \\
PMSM8 & Euler update \eqref{eq:euler} with the parameter network \\
PMSM9 & PGML surrogate trained on PMSM7 (stage D) \\
PMSM10 & FEA machine, bilinear interpolation; stage-D reference \\
Grey-box & Analytical branch of \eqref{eq:residual}, independently fitted to stage-E data \\
PMSM11 & Measured-data residual PGML, learnable heads (stage E) \\
\bottomrule
\end{tabularx}
\end{table}
\section{Proposed Physics-Guided Machine Learning Real-Time Emulator}\label{sec:pgml}
\begin{figure}[tb]
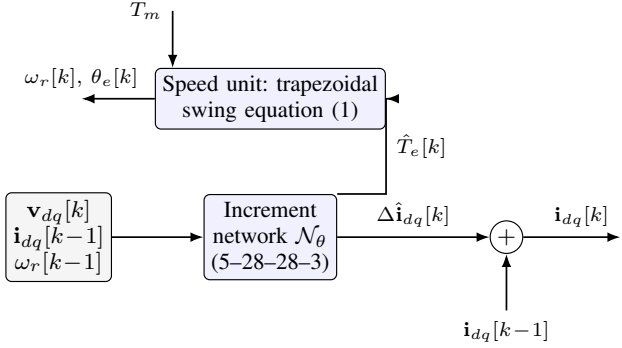

	{\centering
		\resizebox{0.92\columnwidth}{!}{\FigPGML}\par}
	\caption{PGML PMSM model. The increment network predicts the current increment and the torque of one step from the voltages, currents, and speed; speed and angle follow \eqref{eq:linear}. \label{PINN_nonlinearPMSM}}
\end{figure}
\rev{Physics guidance has three forms: analytical one-step labels on sampled states (stages A--C), initialization from the stage-A network when only an FEA-calibrated model exists (stage D), and an analytical branch in the executed step (PMSM11).}

\subsection{Current-Increment Network}
The surrogate of Fig.~\ref{PINN_nonlinearPMSM} advances the currents of one machine with a network $\mathcal{N}_{\theta}$:
\begin{equation}\label{eq:increment}
\begin{aligned}
\big[\Delta\hat{\mathbf i}_{dq}[k],\,\hat T_e[k]\big] &= \mathcal{N}_{\theta}\big([\mathbf v_{dq}[k],\,\mathbf i_{dq}[k\!-\!1],\,\omega_r[k\!-\!1]]\big),\\
\mathbf i_{dq}[k] &= \mathbf i_{dq}[k-1] + \Delta\hat{\mathbf i}_{dq}[k].
\end{aligned}
\end{equation}
The 5--28--28--3 feedforward network has 1067 parameters, a linear first layer, a ReLU second layer, a linear output, and z-score standardized inputs and outputs using training-split statistics. Applying \eqref{eq:increment} recursively makes the currents its dynamic states. PMSM3 adds damper currents $i_{kd}$ and $i_{kq}$ and their increments (7--30--30--5, 1325 parameters). PMSM9 uses 36 neurons per hidden layer (5--36--36--3, 1659 parameters) for the FEA-calibrated machine. On the FPGA, input standardization and the linear first layer are merged into the second layer, leaving one linear--ReLU layer and the output layer.

PMSM8 provides an analytical comparison in stage D: it executes \eqref{eq:euler} with current-dependent parameters supplied by $\mathcal{P}_{\psi}$ (2--12--12--3, 231 parameters),
\begin{equation}\label{eq:paranet}
\mathbf{p}[k] = \alpha\,\mathbf{p}[k-1] + (1-\alpha)\,\mathcal{P}_{\psi}\big(\mathrm{sat}(i_d[k]),\mathrm{sat}(i_q[k])\big),
\end{equation}
where $\mathbf{p}=[L_d,L_q,\lambda_{pm}]^{\mathsf T}$ enters the flux--current relation of \eqref{eq:linear}, $\mathrm{sat}(\cdot)$ limits each input to $\pm2.5$~p.u., and $\alpha=0.97$ (a 33-$\mu$s lag) avoids an algebraic loop; $\mathbf p[k]$ is used at step $k+1$.

\subsection{PMSM11: Residual Form for Measured Data}\label{sec:residual}
For measured drives, PMSM11 combines an analytical current step with a learned residual and physical parameters. The analytical branch supplies the nominal dynamics, while the residual represents effects observed through the drive's voltage-command interface. For a magnetically linear machine, such as the bench drive of Section~\ref{sec:bench}, the complex current $\underline{i}=i_d+ji_q$ evolves in SI units with step $T_s$ as
\begin{equation}\label{eq:residual}
\begin{aligned}
\underline{i}[k+1] &= E\,\underline{i}[k] + \frac{1-E}{R+j\omega_e L}\,\underline{u}[k] + s\,\mathcal{N}_{\theta}\big(\mathbf{x}[k]\big),\\
\underline{u}[k] &= (1-a)\,\underline{v}^*[k] + a\,\underline{v}^*[k-1] - V_{dt}\frac{\underline{i}[k]}{|\underline{i}[k]|} - j\omega_e\underline{\psi},
\end{aligned}
\end{equation}
where $E=\exp(-(R/L+j\omega_e)T_s)$ is the exact zero-order-hold solution, $\underline{v}^*$ is the voltage command of the drive controller, $a$ a fractional command delay, $V_{dt}$ the inverter drop (with $|\underline i|$ replaced by $\sqrt{|\underline i|^2+\varepsilon^2}$, $\varepsilon=0.2$~A), and $\underline{\psi}$ the PM flux linkage with a constant frame offset. The heads $\{R, L, \underline{\psi}, V_{dt}, a\}$ are learnable. PMSM11 uses a 7--28--28--2 network with tanh activations, a zero-initialized output scaled by $s=0.05$~A, and the input $\mathbf{x}=[v_d^*,v_q^*,i_d,i_q,\omega_e,\cos6\theta_e,\sin6\theta_e]^{\mathsf T}$, whose sixth-harmonic terms represent dead-time and flux harmonics. An independent fit of the analytical branch provides the grey-box baseline (Section~\ref{sec:bench_prediction}). The command-referred $R$ and $L$ absorb inverter effects and serve as drift indicators.

\subsection{Training}
\textit{Increment network.} In stages A--C, $\mathcal N_\theta$ uses a data term and a physics term,
\begin{equation}\label{eq:pretrain}
\begin{aligned}
\mathcal{L}_{\text{data}} &= \tfrac{1}{N_d n_y}\textstyle\sum_{n}\big\|\hat{\mathbf y}(\mathbf{x}_n)-\mathbf y^{\text{ref}}_{n}\big\|^2,\\
\mathcal{L}_{\text{phys}} &= \tfrac{1}{N_c n_y}\textstyle\sum_{c}\big\|\hat{\mathbf y}(\mathbf{x}_c)-\mathbf{g}(\mathbf{x}_c)\big\|^2,\\
\mathcal L &= \frac{N_d\mathcal L_{\text{data}}+N_c\mathcal L_{\text{phys}}}{N_d+N_c},
\end{aligned}
\end{equation}
where $\mathbf y=[\Delta\mathbf i_{dq},T_e]$ is standardized and has $n_y$ components (five with dampers), $\mathbf x_n$ are referencer samples, and $\mathbf x_c$ are collocation states drawn uniformly within three standard deviations of each input mean. The analytical one-step map $\mathbf g$ uses the same target standardization. In stages A--C, the data are $9\times10^4$ samples (one per 100 steps) of the 9-s closed-loop run of the referencer and $6\times10^4$ collocation states; the network is trained with Adam \cite{RNNtraining} (batch 256, 300 epochs, learning rate decaying from $10^{-3}$ to $10^{-5}$ by cosine annealing; from $10^{-4}$ when retraining in stage C). \rev{Simulation samples use a 90/10 training/validation split.} In stage D, the network is initialized from stage A, widened to 36 neurons, and trained with $\mathcal L_{\mathrm{data}}$ on $9\times10^4$ PMSM7 samples. The stage-A network follows PMSM1 with an average NMAE of 0.11\%.

\textit{Identified parameters.} PMSM5 uses $L_d$, $L_q$, and $\lambda_{pm}$ identified against PMSM4,
\begin{equation}\label{eq:adapt}
\mathcal{L}_{\text{adapt}}=\tfrac{1}{2N}\textstyle\sum_{n}\big\|\hat{\mathbf{i}}_{dq}\big(\bm{\lambda}_{dq,n};\bm{\theta}_p\big)-\mathbf{i}^{\text{ref}}_{dq,n}\big\|^2,
\end{equation}
where $\hat{\mathbf i}_{dq}$ inverts the flux--current relation of \eqref{eq:linear} with $\bm\theta_p=[L_d,L_q,\lambda_{pm}]$, initialized at $(0.689,\,0.689,\,1.0)$~p.u., over referencer samples of PMSM4, one every 100 steps of its closed-loop run ($N=9\times10^4$; 100 epochs of two full-batch Adam steps, learning rate 0.01).

\textit{Parameter network.} $\mathcal{P}_{\psi}$ is fitted to the gridded map, $\mathcal{L}_{\text{map}}=\frac1M\sum_j\|\mathcal{P}_{\psi}(i_{d,j},i_{q,j})-\mathbf{p}_j\|^2$, with $M=200$ points drawn per epoch in the disk of radius 2.5~p.u.\ and labeled with the nearest grid node (1000 epochs, learning rate $10^{-3}$).

\section{Proposed IIL Architecture and Model Updates}\label{sec:iil}
\begin{figure}[t]
	{\centering
		\includegraphics[width=0.88\columnwidth]{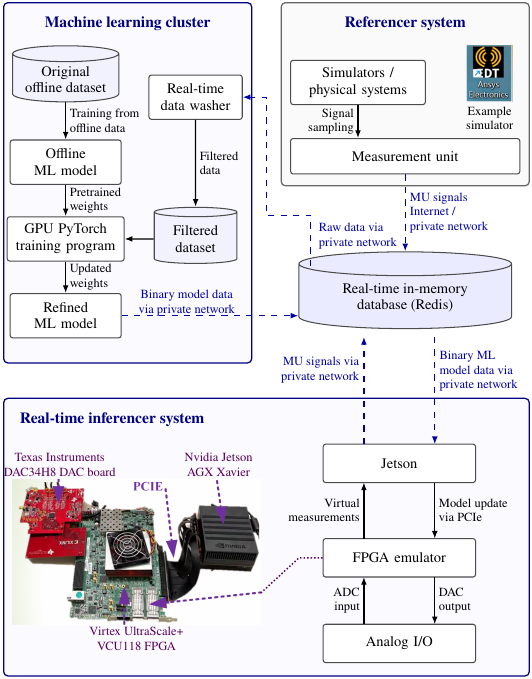}\par}
	\caption{IIL architecture and hardware. The machine learning cluster combines offline pretraining with filtered reference data. The referencer and inferencer exchange measurements and model data through the database; the Jetson releases model updates to the FPGA through PCIe. Dashed blue: network communication; solid black: local data flow; dotted purple: hardware identification. \label{007IILHW}}
\end{figure}

\subsection{IIL Hardware Platform}
The IIL system (Fig.~\ref{007IILHW}) has three components. The \textbf{machine learning cluster} pretrains a baseline model on offline data, filters incoming reference samples, and refines candidate models by GPU training. The \textbf{referencer} acquires trajectories from a simulator or measurements from a physical drive through a measurement unit. A shared in-memory database exchanges these samples, emulator responses, and binary model data over private networks; the referencer can also connect through the Internet. The \textbf{real-time inferencer} integrates a Jetson AGX Xavier, a VCU118 (XCVU9P) FPGA, and a TI DAC34H8 board. PCIe carries model updates and virtual measurements between the Jetson and FPGA, while analog I/O connects the emulator to external hardware. Training and data exchange proceed alongside fixed-step emulation, with the FPGA executing the last accepted model until a validated update is released. A database exchange takes about 100~ms\rev{; it carries only samples and model updates, outside the fixed-step loop}. The FPGA holds preset network structures with writable weights and biases, into which the Jetson writes the folded coefficients (fewer than 1200 values) over PCIe; an update requires no resynthesis or reconfiguration, and its transfer is brief compared with training.

\subsection{Staged Training and Retraining Procedure}
\begin{figure}[!t]
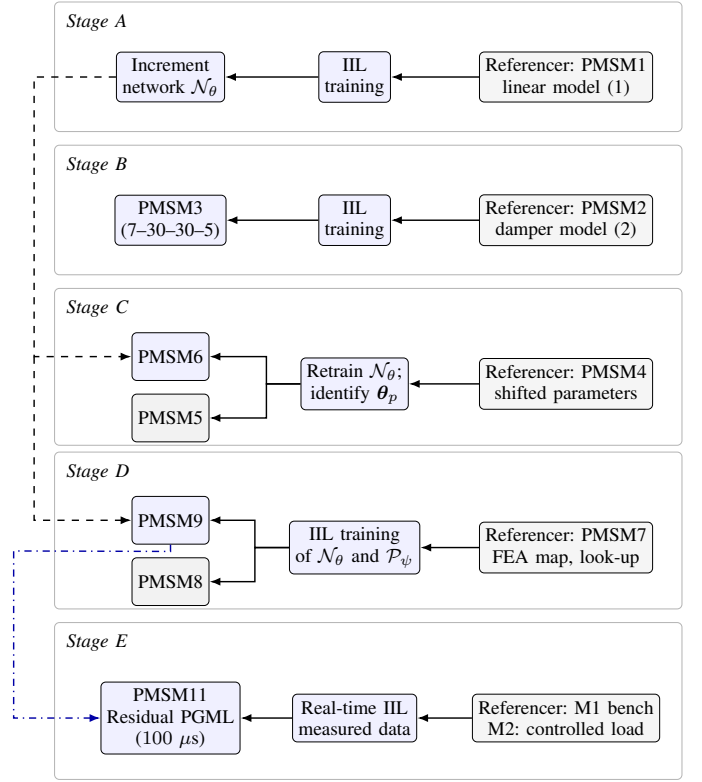

	{\centering
		\FigStages\par}
	\caption{Staged PGML model development. Dashed: initialization from the stage-A weights; blue dash-dotted: structural extension to PMSM11, independently trained on bench data. Gray boxes: analytical models. \label{IIL_PMSM}}
\end{figure}
Updates are accepted when validation loss decreases and physical parameters remain within bounds. Fig.~\ref{IIL_PMSM} shows five stages: A, pretraining on PMSM1; B, training PMSM3 against PMSM2 with dampers; C, retraining PMSM6 against PMSM4 and identifying parameters for the analytical PMSM5; D, training PMSM9 on PMSM7 and fitting the parameter network of PMSM8 to the same map, both evaluated against PMSM10; and E, training PMSM11 on measured M1 data with M2 as a controlled load. C and D initialize from stage A and retrain \rev{the whole network} for their referencers. PMSM11 uses an independently trained current residual and physical heads at 100~$\mu$s\rev{, and only its heads are updated online} (Section~\ref{sec:bench}).

\begin{figure}[t]
	{\centering
		\includegraphics[width=0.88\columnwidth]{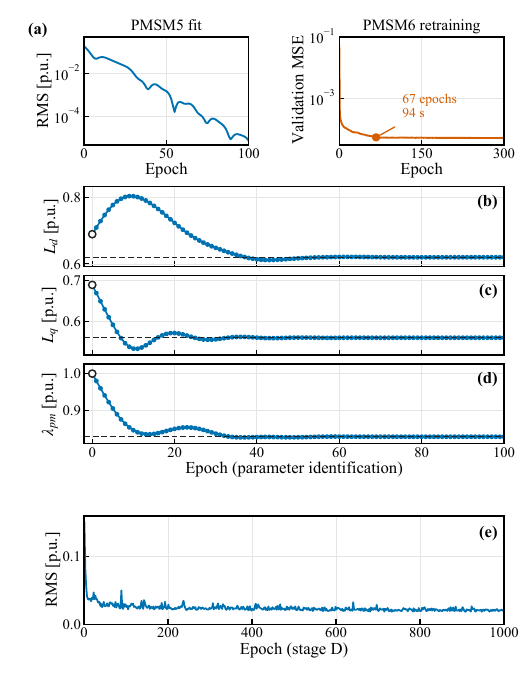}\par}
	\caption{Model updates from training logs. (a) PMSM5 current-fit RMS (left) and PMSM6 validation mean squared error in standardized outputs (right), with epoch 67 (94~s) marked. (b)--(d) Identified $L_d$, $L_q$, and $\lambda_{pm}$ (circles: initialization; dashed: PMSM4). (e) Stage-D parameter-network test RMS error. \label{IIL_update3}}
\end{figure}

Fig.~\ref{IIL_update3} shows the training cost of each update. PMSM6 retraining reaches within 5\% of its final validation loss after 67 epochs, taking 94~s on one desktop GPU, and completes 300 epochs in 402~s. Initialization from stage A roughly halves this effort; from random weights, 126 epochs are needed. The analytical comparator PMSM5 reaches $(0.6200,\,0.5600,\,0.8300)$~p.u.\ in 100 epochs (1.48~s on one CPU core), with a current-fit RMS of $7.71\times10^{-6}$~p.u. The stage-D parameter-network test RMS decreases from about 0.15 to 0.02~p.u.\ over 1000 epochs.

\subsection{FPGA Implementation}
\begin{table}[t]
\centering
\caption{FPGA Resources and Latency per Time Step on the VCU118 (XCVU9P FPGA) at 100~MHz}\label{tab:007resources}
\footnotesize
\renewcommand{\arraystretch}{1.1}
\setlength{\tabcolsep}{2.0pt}
\begin{tabular}{lcccccc}
\toprule
\textbf{Model} & \textbf{BRAM} & \textbf{DSP} & \textbf{FF} & \textbf{LUT} & \textbf{Cycles} & \textbf{$\mu$s} \\
\midrule
\multicolumn{7}{l}{\textit{1-$\mu$s simulation kernels, fixed point}} \\
\rev{Analytical (PMSM5)} & \rev{0} & \rev{12} & \rev{0.92k} & \rev{3.0k} & \rev{75} & \rev{0.75} \\
\rev{Anal.\ + param.\ net (PMSM8)} & \rev{0} & \rev{54} & \rev{8.6k} & \rev{17.2k} & \rev{84} & \rev{0.84} \\
\rev{LUT, nearest (PMSM7)} & \rev{5} & \rev{17} & \rev{8.7k} & \rev{9.2k} & \rev{114} & \rev{1.14} \\
\rev{LUT, bilinear (PMSM10)} & \rev{5} & \rev{18} & \rev{9.5k} & \rev{10.3k} & \rev{141} & \rev{1.41} \\
PGML, dampers (PMSM3) & \rev{0} & \rev{103} & \rev{7.7k} & \rev{13.8k} & \rev{33} & \rev{0.33} \\
PGML (PMSM6) & 0 & \rev{85} & \rev{6.6k} & \rev{11.4k} & \rev{30} & \rev{0.30} \\
PGML, FEA machine (PMSM9) & \rev{0} & \rev{101} & \rev{8.0k} & \rev{13.0k} & \rev{34} & \rev{0.34} \\
\multicolumn{7}{l}{\textit{100-$\mu$s bench kernels, single-precision floating point}} \\
\rev{Grey-box} & \rev{0} & \rev{8} & \rev{1.9k} & \rev{1.9k} & \rev{82} & \rev{0.82} \\
PMSM11 & 5 & 203 & \rev{30.5k} & \rev{25.6k} & \rev{234} & \rev{2.34} \\
\textbf{Available} & 4320 & 6840 & 2364k & 1182k & {} & {} \\
\bottomrule
\multicolumn{7}{p{0.95\columnwidth}}{\scriptsize Per-step $dq$ electrical kernel; the voltage and current transforms, mechanics, and I/O are common to all rows and add at most 10 cycles per step. BRAM in 18-Kb blocks. Measured on the FPGA.}
\end{tabular}
\end{table}

Table~\ref{tab:007resources} shows that the PGML kernels use \rev{30--34} cycles and \rev{85--103} DSPs. \rev{The analytical PMSM5 takes 75 cycles and PMSM8 84, including its parameter network, filter, and division.} With the common transforms, mechanics, and I/O, a complete PGML step takes at most \rev{44} cycles (\rev{0.44}~$\mu$s), \rev{and the PGML and analytical kernels fit the 1-$\mu$s step, whereas the look-up kernels PMSM7 and PMSM10 take 1.14--1.41~$\mu$s, comparable to the 1.5--2-$\mu$s steps reported for FEA-based LUT models on a commercial real-time simulator \cite{Li2026}.} The bench kernels take \rev{0.82}~$\mu$s and \rev{2.34}~$\mu$s within the 100-$\mu$s period; PMSM11 computes its analytical and residual branches in parallel. Each representation uses a small fraction of the device: the FEA table has 1443 values, the parameter network 231 parameters, and PMSM9 1659 parameters.
\section{Simulation Validation}\label{sec:results}
All models run in the same closed-loop drive: a two-level inverter on a 540-V dc link with 12-kHz carrier-based PWM and min--max zero-sequence injection, field-oriented control with $i_d^*=-0.5$~p.u., a speed proportional--integral (PI) controller, and current PI controllers with decoupling feedforward. The feedforward uses $L=0.689$~p.u.\ for all models and the PM flux of the model under test. The 9-s sequence holds standstill until 0.1~s; sets $\omega^*=0.5$~p.u.\ with $T_m=0.5$~p.u.; ramps to 0.2~p.u.\ at 1--1.5~s; steps to 1.2~p.u.\ at 2.5~s; ramps to 0.9~p.u.\ at 5.5--6~s; removes the load at 7--8~s; and applies $T_m=0.8$~p.u.\ at 8--9~s, moving the operating point across the nonlinear region of the maps. For a p.u. response $x\in\{i_d,i_q,T_e,\omega_r\}$, the normalized mean absolute error (NMAE) of model $m$ against its reference $r$ is given as:
\begin{equation}
\operatorname{NMAE}_{x}^{(m,r)}=\frac{100}{N}\textstyle\sum_{k=1}^{N}\left|x_m[k]-x_r[k]\right|,
\label{eq:mpae}
\end{equation}
in percent of the base value, which avoids division by the zero-crossing $i_d$. The responses in Table~\ref{tab:mpae_quantitative} are recorded from the FPGA execution of each model.

\begin{table}[t]
\caption{NMAE (\%) of P.U. Responses Over the 9-s Sequence; Darker Cells Indicate Larger Values}
\label{tab:mpae_quantitative}
\centering
\scriptsize
\setlength{\tabcolsep}{2.0pt}
\begin{tabular}{c c c c c c c}
\toprule
Model & Ref. & $i_d$ & $i_q$ & $T_e$ & $\omega_r$ & Avg. \\
\midrule
\multicolumn{7}{l}{\textit{Base-machine models vs.\ FEA machine (before transfer)}} \\
PMSM1 & PMSM10 & \MPAEcell{53}{4.813} & \MPAEcell{85}{7.741} & \MPAEcell{25}{2.283} & \MPAEcell{10}{0.953} & \MPAEcell{43}{3.947} \\
PMSM2 & PMSM10 & \MPAEcell{55}{5.028} & \MPAEcell{84}{7.688} & \MPAEcell{33}{3.005} & \MPAEcell{10}{0.880} & \MPAEcell{46}{4.150} \\
PMSM3 & PMSM10  & \MPAEcell{55}{5.019} & \MPAEcell{86}{7.800} & \MPAEcell{35}{3.142} & \MPAEcell{10}{0.896} & \MPAEcell{46}{4.214} \\
\multicolumn{7}{l}{\textit{Stage B: damper training}} \\
PMSM3 & PMSM2   & \MPAEcell{12}{1.069} & \MPAEcell{9}{0.798} & \MPAEcell{8}{0.754} & \MPAEcell{5}{0.033} & \MPAEcell{7}{0.663} \\
\multicolumn{7}{l}{\textit{Stage C: parameter shift}} \\
PMSM5 & PMSM4   & \MPAEcell{5}{0.082} & \MPAEcell{5}{0.071} & \MPAEcell{5}{0.058} & \MPAEcell{5}{0.001} & \MPAEcell{5}{0.053} \\
PMSM6 & PMSM4   & \MPAEcell{5}{0.267} & \MPAEcell{5}{0.273} & \MPAEcell{5}{0.191} & \MPAEcell{5}{0.016} & \MPAEcell{5}{0.187} \\
\multicolumn{7}{l}{\textit{Stage D: FEA machine (after transfer)}} \\
PMSM7 & PMSM10 & \MPAEcell{6}{0.546} & \MPAEcell{5}{0.411} & \MPAEcell{5}{0.383} & \MPAEcell{5}{0.021} & \MPAEcell{5}{0.340} \\
PMSM8 & PMSM10 & \MPAEcell{7}{0.662} & \MPAEcell{12}{1.064} & \MPAEcell{6}{0.586} & \MPAEcell{5}{0.072} & \MPAEcell{7}{0.596} \\
PMSM9 & PMSM10  & \MPAEcell{12}{1.066} & \MPAEcell{9}{0.849} & \MPAEcell{7}{0.644} & \MPAEcell{5}{0.129} & \MPAEcell{7}{0.672} \\
\bottomrule
\end{tabular}
\end{table}

\begin{figure}[t]
	{\centering
		\includegraphics[width=0.88\columnwidth]{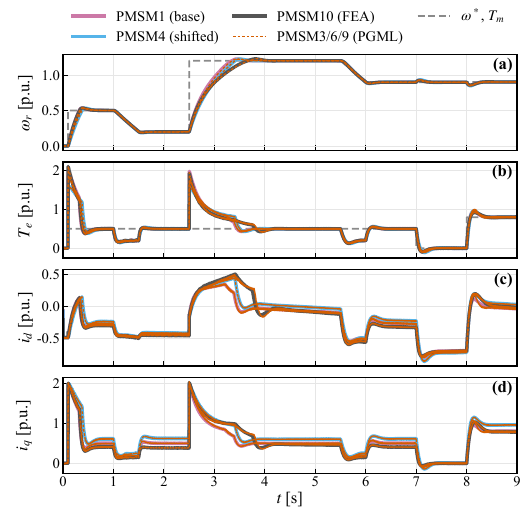}\par}
	\caption{Closed-loop responses of the base machine (PMSM1), shifted machine (PMSM4), and FEA machine (PMSM10), overlaid by the PGML surrogates PMSM3, PMSM6, and PMSM9 (orange dashed): (a) $\omega_r$, (b) $T_e$, (c) $i_d$, and (d) $i_q$, as 250-$\mu$s averages; dashed gray: $\omega^*$ and $T_m$. Fig.~\ref{TS_results} shows deviations from the stage references. \label{Current_results}}
\end{figure}

\begin{figure*}[t]
	{\centering
		\includegraphics[width=0.95\textwidth]{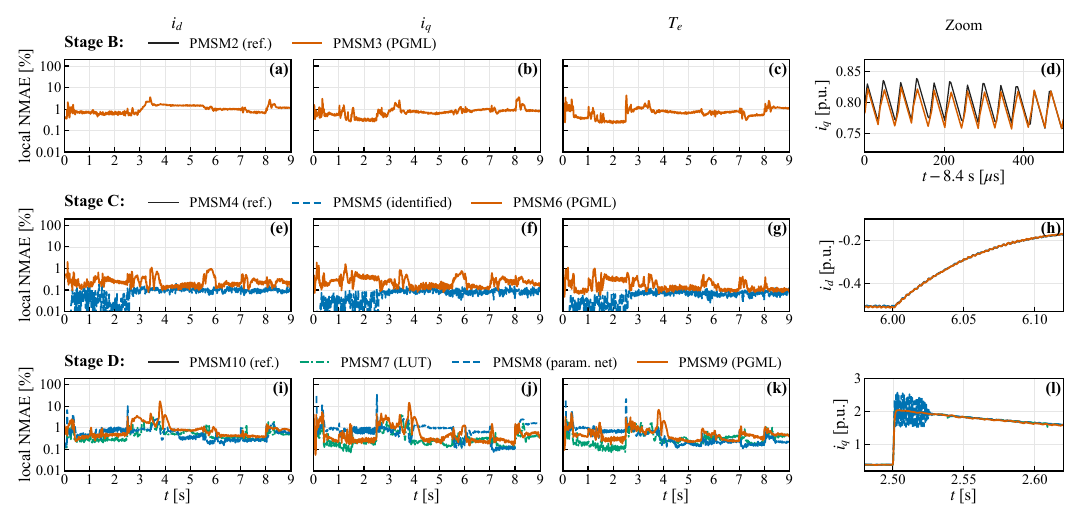}\par}
	\caption{Deviation of each surrogate from its stage reference (black): stage B (top row), stage C (middle), stage D (bottom). (a)--(c), (e)--(g), (i)--(k) Local NMAE of $i_d$, $i_q$, and $T_e$ over 10-ms windows; time averages match the corresponding columns of Table~\ref{tab:mpae_quantitative}. (d) $i_q$ of PMSM2 and PMSM3 at every 1-$\mu$s step at 8.4~s. (h) $i_d$ at the end of the speed ramp at 6~s and (l) $i_q$ after the step at 2.5~s, as 250-$\mu$s averages. \label{TS_results}}
\end{figure*}

\subsection{Closed-Loop Responses}
Fig.~\ref{Current_results} shows the responses of the three machines under the same excitation. Table~\ref{tab:mpae_quantitative} evaluates each model against its stage reference, and Fig.~\ref{TS_results} resolves the deviations over time. The steady-state $i_q$ of PMSM4 exceeds that of the other machines due to its reduced PM flux, and PMSM1 and PMSM10 differ most in the high-speed interval from 2.5 to 5.5~s, where the FEA machine operates deepest in its nonlinear region. Before stage D, the base-machine models PMSM1--PMSM3 deviate from the FEA machine by 3.95\%--4.21\% on average, which quantifies the model-form gap closed by stage D.

\textit{Stage B.} PMSM3 represents the damper dynamics with an average NMAE of 0.66\%. Its $i_q$ switching ripple near 8.4~s is 0.0676~p.u.\ peak to peak, compared with 0.0811~p.u.\ for PMSM2; the mean differs by $-0.00842$~p.u.\ (Fig.~\ref{TS_results}(d)).

\textit{Stage C.} Retraining PMSM6 for the shifted machine gives an average NMAE of 0.19\% with the same current-increment form as stage A. The identified $L_d$, $L_q$, and $\lambda_{pm}$ are 0.6200, 0.5600, and 0.8300~p.u., within 0.01\% of PMSM4, and the analytical PMSM5 reaches 0.05\%. Because the simulation referencers are these discretized equations, PMSM5 is the reference point of this stage, and PMSM6 lies 0.13 percentage points above it. The physics term generalizes the learned step beyond the recorded trajectory: on $10^4$ fresh states its one-step error is 0.71\%, against 3.87\% for a network trained on $\mathcal L_{\mathrm{data}}$ alone, and its currents decay under zero input. PMSM6's largest local electrical NMAE is 2.00\% in $i_d$ at 0.125~s. After the ramp ends at 6~s, its $i_d=-0.35$~p.u.\ crossing follows PMSM4 by 0.70~ms (Fig.~\ref{TS_results}(h)).

\textit{Stage D.} PMSM8 and PMSM9 follow the FEA machine with 0.60\% and 0.67\%, i.e., 14\%--17\% of the pre-transfer errors. PMSM7, from which both learn, deviates from PMSM10 by 0.34\% through the nearest-neighbor quantization of the map, which sets the reference level of this stage. PMSM8 combines the analytical step with the parameter network and filter \eqref{eq:paranet}, whereas PMSM9 encodes the FEA-calibrated dynamics in a single network learned from referencer data. Following the steps at 0.1 and 2.5~s, the 250-$\mu$s-averaged $i_q$ peaks are respectively 2.52 and 2.57~p.u.\ for PMSM8, 2.01 and 2.04~p.u.\ for PMSM9, and 2.00 and 2.02~p.u.\ for PMSM10. PMSM8 oscillates for about 23 and 27~ms (detrended ripple above 0.1~p.u.), whereas the detrended ripple of PMSM9 stays below 0.015~p.u., as for PMSM10 (Fig.~\ref{TS_results}(l)). This oscillation stems from the loop through the parameter network and lasts 34--37~ms without the filter ($\alpha=0$). For the stage-specific comparisons, speed NMAE remains below 0.13\%, with mechanical integration and speed feedback attenuating the electrical-model differences.

\rev{\textit{Independent profile.} A 3-s profile not used in training, with other speed and load steps in the forward-motoring range of the 9-s sequence, keeps all models stable, with average NMAEs of 0.51\% for PMSM6 and 1.25\% for PMSM9. The learned steps deviate most during braking after a speed step, where $i_q$ leaves the range of the training trajectory; IIL extends such ranges by retraining on reference data that cover them.}

\subsection{Fault Transients and Parameter Maps}
\begin{figure}[t]
    {\centering
        \includegraphics[width=1\columnwidth]{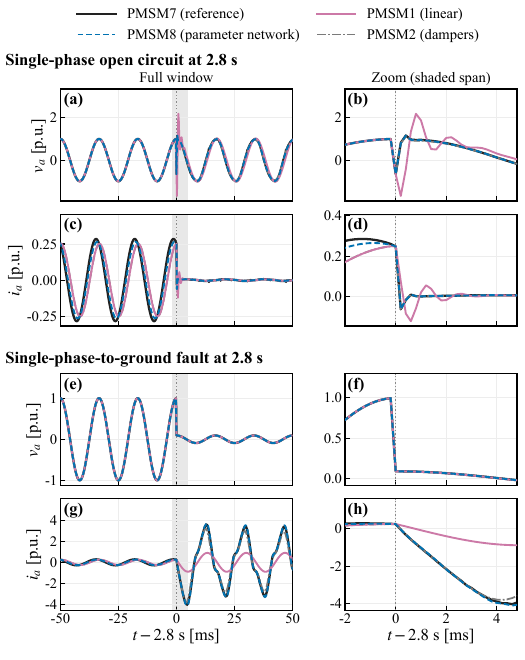}\par}
    \caption{Phase-A voltage and current under faults at 2.8~s: (a), (b) $v_a$ and (c), (d) $i_a$ for a single-phase open circuit; (e)--(h) the same for a phase-to-ground fault. Right: zoom into the shaded span; PMSM7 (black) is the reference. \label{Fault_re}}
\end{figure}
\begin{figure}[t]
	{\centering
		\includegraphics[width=0.88\columnwidth]{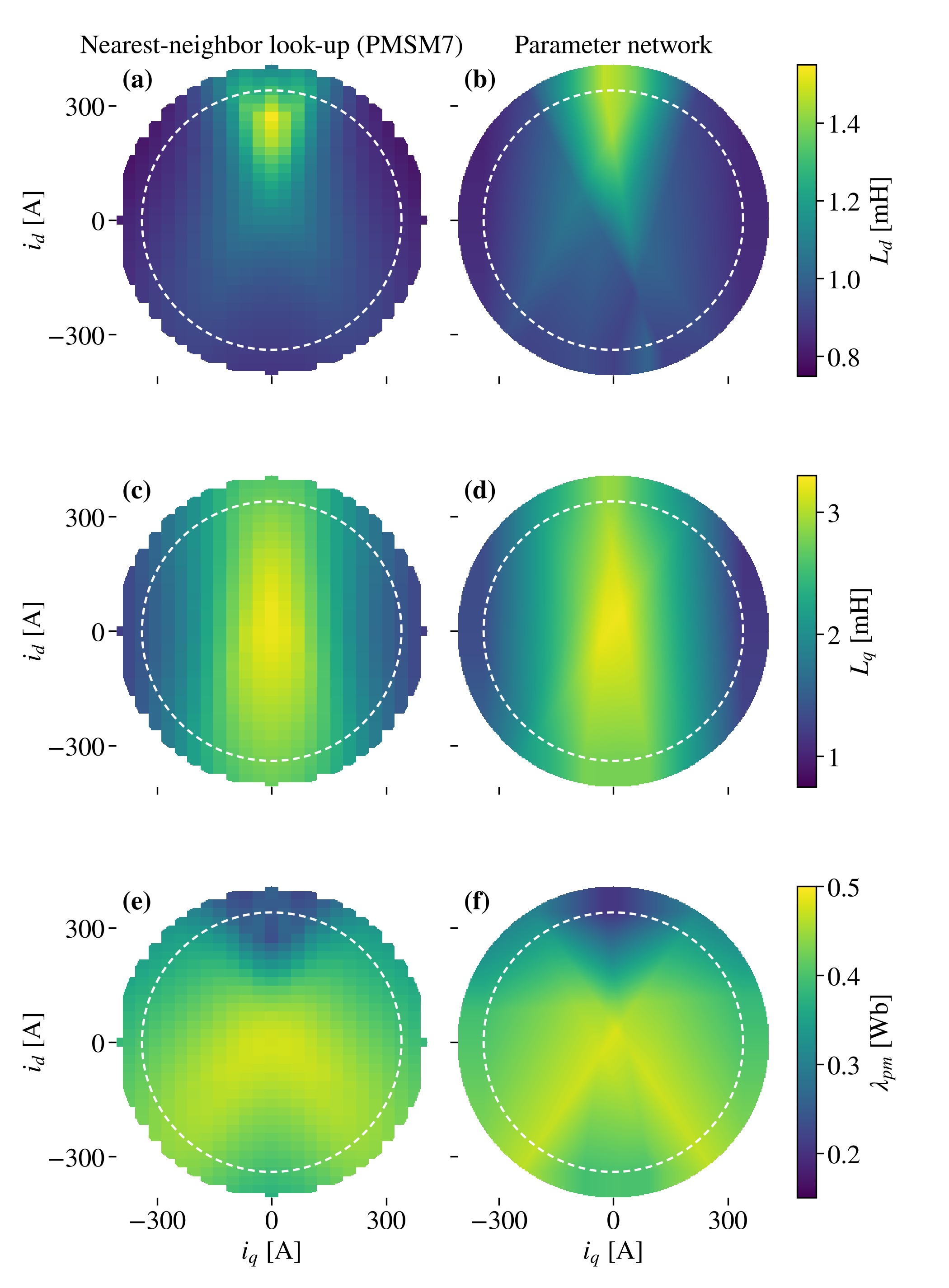}\par}
	\caption{FEA parameter maps over the current plane: nearest-neighbor look-up executed by PMSM7 (left) and parameter-network output (right) for (a), (b) $L_d$; (c), (d) $L_q$; (e), (f) $\lambda_{pm}$. The dashed circle marks 2.5~p.u. \label{007LUT_PINN_results}}
\end{figure}
In the fault study (Fig.~\ref{Fault_re}), the machine is fed at a light load of 0.2~p.u.\ from an ideal 60-Hz, 1-p.u.\ source through series, ground, and snubber branches, and PMSM7 and PMSM8 include the damper circuits. Without dampers, PMSM1 rings after the open circuit and underestimates the ground-fault current (0.91 against 3.81~p.u.\ for PMSM2). PMSM8 follows PMSM7 through both faults, with a largest phase-current difference of 0.19~p.u., under 5\% of the 4.04-p.u.\ peak; the ground-fault current extends beyond the FEA range, where both extrapolate the same map. Fig.~\ref{007LUT_PINN_results} compares the executed maps. The network reproduces the asymmetry of $L_d$, the ridge of $L_q$, and the reduction of $\lambda_{pm}$, and is continuous where the look-up is piecewise constant. Against the 95 FEA points within 2.5~p.u., its RMS errors are 0.010, 0.013, and 0.015~p.u.\ for $L_d$, $L_q$, and $\lambda_{pm}$ (3.3\%, 1.6\%, and 1.5\% of the means).
\section{Experimental Validation on a 200-W Drive}\label{sec:bench}
The 200-W setup evaluates the IIL emulator on a physical drive; results use the controller base in Table~\ref{tab:bench}.

\subsection{Test Bench, Identification, and Closed-Loop Twin}
\begin{figure*}[t]
  \centering
  \includegraphics[width=0.92\textwidth,height=0.25\textheight,keepaspectratio]{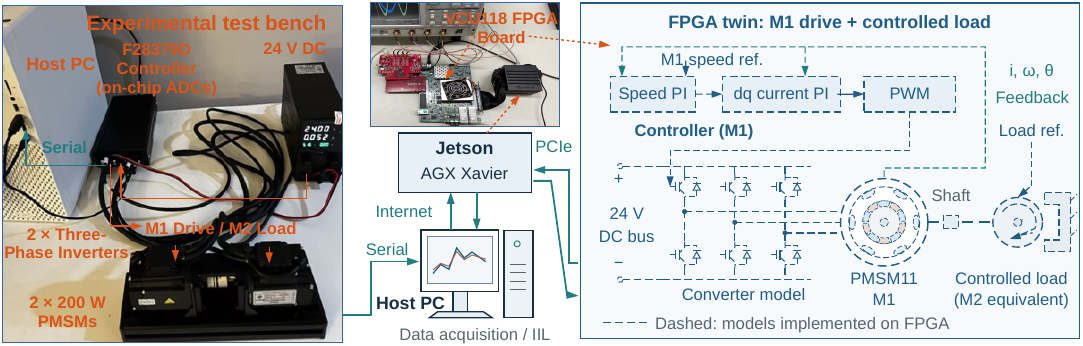}
  \caption{Proposed real-time IIL test bench for PMSM11. M1 is the 200-W PMSM under study; the coaxially coupled M2 supplies a controllable load. The F28379D streams measurements to the host PC via serial communication. The host PC exchanges data and model updates with the Jetson over the Internet, and the Jetson connects to the FPGA through PCIe. The FPGA implements the M1 controller, converter, and PMSM11 models, with M2 represented by an equivalent controlled load-torque model. Dashed outlines denote FPGA models.}
  \label{fig:bench}
\end{figure*}

\begin{table}[t]
\centering
\footnotesize
\caption{Test Bench and Parameters of M1}
\label{tab:bench}
\renewcommand{\arraystretch}{1.05}
\setlength{\tabcolsep}{3pt}
\begin{tabular}{ll}
\toprule
\textbf{Quantity} & \textbf{Value} \\ \midrule
Machine & SPM, 200~W, $p=5$, 0.64~N$\cdot$m rated \\
DC link; PWM; current/speed loop & 24~V; 20~kHz; 100~$\mu$s/1~ms \\
Controller base current, voltage & 21.43~A, $24/\sqrt3$~V \\ \midrule
\rowcolor[gray]{0.95} \multicolumn{2}{l}{\textit{Identified from terminal voltages}} \\
Resistance per phase, incl. leads & 72.1 (70--76)~m$\Omega$ \\
Inductance ($L_d$, $L_q$ fitted apart) & 0.121 (0.117, 0.126)~mH \\
PM flux linkage; coupled inertia & 5.29~mWb; $6.64\times10^{-5}$~kg$\cdot$m$^2$ \\ \midrule
\rowcolor[gray]{0.95} \multicolumn{2}{l}{\textit{Command-referred heads of \eqref{eq:residual}, PGML (grey-box)}} \\
$R$; $L$ & 90.8 (88.9)~m$\Omega$; 0.287 (0.182)~mH \\
$|\underline\psi|$; $V_{dt}$ & 5.10 (5.18)~mWb; 0.337 (0.278)~V \\
$a$; frame offset of $\underline\psi$ & 0.046 (0.065); $5.2^\circ$ ($7.0^\circ$) \\
\bottomrule
\end{tabular}
\end{table}

The bench (Fig.~\ref{fig:bench}) consists of two PMSMs on a common shaft, fed from a 24-V dc link and controlled by a TI TMS320F28379D whose code is generated from Simulink: M1, the machine under study, is speed-controlled, while M2 uses its current loop to impose a controllable opposing shaft torque. During operation, the controller streams to a host PC every 50-$\mu$s sample of the phase currents, terminal voltages, and encoder position, together with the $dq$ voltage commands of every 100-$\mu$s control period; valid received samples retain their acquisition times. Quantities use the on-board sensors at nominal gains, and winding temperature is inferred from resistance. The bench operates between 300 and 2000~rpm with load currents up to 6~A. In this range M1 is magnetically linear; the bench thus exercises inverter nonlinearities and temperature-dependent parameters rather than saturation.

The Table~\ref{tab:bench} parameters were identified from 517~s of 20-kHz data by integral regression of the $\alpha\beta$ voltage equation on the terminal voltages; on an independent run, it predicts the current vector from measured voltages with a relative RMS error of 9.8\%. Separate $d$- and $q$-axis fits differ by 7\% with little gain on independent data, so the machine is treated as magnetically linear. The command-referred heads exceed the machine values as they include the inverter drop and delay.

The closed-loop twin combines the M1 controller, identified machine, converter drop and delay, shaft inertia, and drag with an equivalent controlled M2 load torque. Fixed before validation, it follows speed and load references on six subsequently recorded runs: random steps, a ship-voyage profile with propeller-type load, and a 2000-rpm, 3.5-A corner case, each run twice. Scored durations are listed in Table~\ref{tab:bench_holdout}. Relative RMS errors are 0.72\%--1.12\% in speed, 8.73\%--16.57\% in $i_q$, and 1.05\%--2.94\% in the $q$-axis voltage command. These closed-loop results use the analytical M1 model; PMSM11 is evaluated below using the measured angle.

\subsection{PMSM11 on Measured Data}\label{sec:bench_prediction}
\begin{table}[t]
\caption{NMAE (\% of 21.43~A) of Open-Loop Current Prediction on the Six Bench Test Runs}
\label{tab:bench_holdout}
\centering
\scriptsize
\setlength{\tabcolsep}{2.6pt}
\begin{tabular}{l c ccc ccc}
\toprule
& & \multicolumn{3}{c}{$i_d$} & \multicolumn{3}{c}{$i_q$} \\
\cmidrule(lr){3-5}\cmidrule(lr){6-8}
Test run & Scored & Anal. & Grey & PMSM11 & Anal. & Grey & PMSM11 \\
\midrule
Random, 1st & 268~s & 11.51 & 1.31 & \textbf{0.59} & 15.43 & 1.65 & \textbf{0.89} \\
Random, 2nd & 267~s & 11.62 & 1.35 & \textbf{0.58} & 15.03 & 1.60 & \textbf{0.88} \\
Voyage, 1st & 215~s & 13.95 & 1.35 & \textbf{0.52} & 18.00 & 1.44 & \textbf{1.02} \\
Voyage, 2nd & 220~s & 14.12 & 1.46 & \textbf{0.52} & 17.78 & 1.44 & \textbf{1.16} \\
Corner, 1st & 110~s & 18.04 & 2.76 & \textbf{1.06} & 25.60 & 2.36 & \textbf{1.81} \\
Corner, 2nd & 113~s & 18.09 & 2.94 & \textbf{1.11} & 25.28 & 2.47 & \textbf{1.90} \\
\bottomrule
\end{tabular}
\end{table}

\begin{figure}[t]
	{\centering
		\includegraphics[width=0.95\columnwidth]{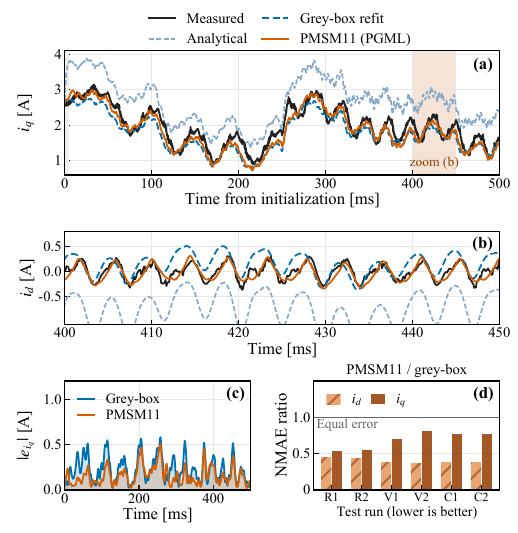}\par}
	\caption{Current prediction on held-out bench data. (a) $i_q$ in a 0.5-s corner-case chunk (2000~rpm, 3.5-A load); (b) $i_d$ in the shaded span. Each model rolls forward from one measured current sample using commanded voltages and measured speed and angle; scoring starts after the gray 5-ms initialization interval. (c) Absolute $i_q$ error, averaged over 5~ms. (d) PMSM11 NMAE relative to the grey-box NMAE on all six runs (R: random steps; V: voyage; C: corner case; Table~\ref{tab:bench_holdout}). \label{fig:bench_holdout}}
\end{figure}

The grey-box model and PMSM11 are fitted at $T_s=100~\mu$s on about 245~s of valid data from nine dedicated excitation runs (current and speed steps, rotating voltage pulses, accelerations, decelerations, and periodic $q$-axis excitation). The grey-box baseline independently optimizes $R$, $L$, $\underline{\psi}$, $V_{dt}$, and $a$ in the analytical branch of \eqref{eq:residual} with Adam, minimizing the mean squared current error over 2-ms free runs initialized from one measured sample. PMSM11 initializes its physical parameters from this fit and jointly optimizes them with the residual network using the same loss. Both models are then refined over 10-ms free runs, with the last 10\% of each run used for checkpoint selection. The analytical baseline uses the identified machine parameters with the inverter drop and one period of delay. The six later runs form the frozen test set.

Across the six test runs, PMSM11 lowers $i_d$ NMAE to 36\%--45\% and $i_q$ NMAE to 54\%--81\% of the grey-box errors (Table~\ref{tab:bench_holdout} and Fig.~\ref{fig:bench_holdout}(d)). Its absolute ranges are 0.52\%--1.11\% and 0.88\%--1.90\%, compared with 1.31\%--2.94\% and 1.44\%--2.47\% for the grey-box refit. Each 0.5-s chunk rolls forward from one measured initial current using commanded voltages and measured speed and angle; 2411 chunks cover 1193~s of valid data. Rolled forward over each entire run of 112--274~s from a single initial sample, PMSM11 \rev{gives} the same errors (e.g., 0.59\% and 0.89\%), i.e., 0.11--0.24~A ($i_d$) and 0.19--0.41~A ($i_q$)\rev{; with a 3-ms electrical time constant, the initial current fades within each chunk}. The terminal-voltage analytical model reaches 11.5\%--25.6\% when driven through the command interface. Relative to measured $i_q$ RMS, errors are 5.9\%--9.4\% for PMSM11, 10.4\%--13.2\% for the grey-box refit, and 93\%--116\% for the analytical model.

\subsection{Online Parameter Tracking During Heating}
\begin{figure}[t]
	{\centering
		\includegraphics[width=0.95\columnwidth]{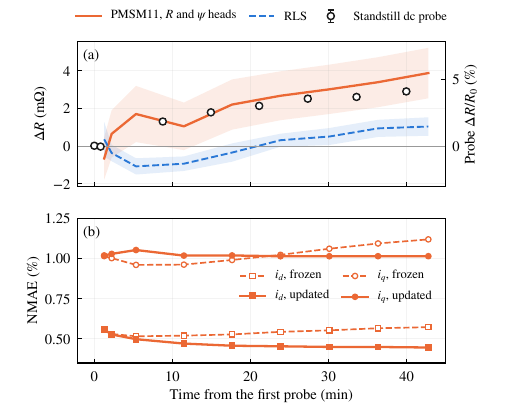}\par}
	\caption{Thermal test with online PMSM11 updates. (a) Change of the command-referred $R$ from jointly updating the $R$ and $\underline\psi$ heads (solid) and from recursive least squares (RLS, dashed), as segment means $\pm1$ standard deviation over 1-s blocks, and the phase-mean winding-resistance change from standstill dc probes (circles, $\pm1$ standard error). (b) Current NMAE per loaded segment before updating, for the frozen (dashed) and updated (solid) model. \label{fig:bench_thermal}}
\end{figure}

\textit{Independent resistance reference.} In the thermal test, M1 runs at 1000~rpm while M2 switches its load current between 2 and 4~A every 2~s, so that the load-dependent resistive voltage separates from the flux-linkage voltage. Every 5~min of loading, a standstill probe injects 2, 4, and 6~A dc into each phase and takes the resistance as the slope of terminal voltage over current; two initial probes agree to 0.27\%. Over 30~min of loading, the resistance rises from 71.28 to 74.18~m$\Omega$ ($+4.1\%$, about 10~$^\circ$C), with standard errors of about 0.13~m$\Omega$.

\textit{Online update from operating data.} The recorded voltage commands, currents, and position are streamed in order to PMSM11. Each 1-s block is first predicted and scored; ten Adam steps then update $\log R$ (learning rate $4\times10^{-3}$) and the in-phase component of $\underline\psi$ ($4\times10^{-4}$) on 256 windows of 10~ms from that block, with the network and the other heads fixed. Each update takes at most 0.44~s on a desktop CPU. Referred to the two initial loaded segments, the $R$ head follows the probe increments with an RMS error of 0.44~m$\Omega$ over the six heating probes, reaching $+3.40$ against $+2.91$~m$\Omega$ at 30~min, while the flux head decreases by 1.3\%\rev{; the RMS error is about 15\% of the 2.91-m$\Omega$ rise}. Recursive least squares (RLS) \rev{estimating $R$ and $\psi$ jointly} on the analytical branch (60-s forgetting horizon) reaches 2.32~m$\Omega$\rev{, so updating the heads inside the executed step, beside the residual, tracks the rise about five times more closely}. Updating both heads keeps the flux decrease out of $R$.

Scored before each update, the updated model keeps its $i_q$ NMAE at 1.01\%--1.05\% and lowers its $i_d$ NMAE from 0.56\% to 0.45\%, whereas the frozen model drifts to 0.57\% and 1.12\% by the last segment (Fig.~\ref{fig:bench_thermal}(b)); the grey-box model stays at 1.32\%--1.43\% and 1.35\%--1.41\%. Because the heads are command-referred, they lump winding, magnet, and inverter changes; in this test, the $R$ head follows the winding resistance.

\subsection{Real-Time Execution}
The FPGA executes PMSM11 at the 100-$\mu$s control period together with the M1 controller and converter models and an equivalent M2 load-torque model (Fig.~\ref{fig:bench}, Table~\ref{tab:007resources}). \rev{The drive samples at up to 20~kHz; the 100-$\mu$s step of PMSM11, a trade-off, matches the current-control period of the voltage commands and keeps the streamed data within the serial-link bandwidth. Each step of the proposed RTDT completes in under 2.5~$\mu$s, a margin of 20 times even at 50-$\mu$s sampling, so each twin step is ready long before the drive delivers its next measurement.} The DAC interface also supports coupling the emulator to external controllers for HIL tests.

\section{Engineering Implications}\label{sec:limits}
\textit{Accuracy and cost.} Where exact equations or maps exist, \rev{the analytical and look-up models are up to 0.33 percentage points more accurate}: PMSM5 reaches 0.05\% against 0.19\% for PMSM6, and PMSM7 0.34\% against 0.67\% for PMSM9. \rev{In exchange,} the same \rev{30--34}-cycle kernel covers dampers and FEA saturation\rev{, well within the 1-$\mu$s step. A network kernel is matrix multiplication on parallel DSP arrays, so resources buy speed directly: the PGML kernels use more DSPs than the analytical and look-up ones and run faster, and one optimized design serves every machine, whereas each equation-based model needs its own pipelining, division, and table trade-off.}

\textit{Model choice.} With a few known parameters, the analytical step is \rev{more accurate}; with an accurate flux map from FEA or characterization tests, a look-up model is the \rev{more accurate choice where its longer step is acceptable,} and is regenerated offline when the machine changes. PGML suits references that consist of recorded operation, as for PMSM11\rev{; its updates range from a few heads adjusted online to the whole executed step retrained on the GPU, without changing the FPGA design}.

\section{Conclusion}\label{sec:conclusion}
HIL tests are only as trustworthy as the machine model inside the RTDT\rev{, which has to follow the installed machine as it changes}. This article presented a physics-guided IIL emulator in which the FPGA executes a learned one-step current model retrained from reference data. A compact current-increment network represents damper dynamics and FEA-calibrated saturation with average NMAE of 0.19\%--0.67\% and measured electrical-kernel latencies of \rev{0.30--0.34}~$\mu$s. On a 200-W drive, the residual PMSM11 runs 20 times faster than real time \rev{even at 20-kHz sampling}, improves both current axes over an independently fitted grey-box model on all six held-out test runs, and, updated online from operating data during 30~min of heating, tracks the probe-measured resistance rise with an RMS error of 0.44~m$\Omega$. Future work will test the RTDT on a saturated machine, trigger updates automatically over full thermal cycles on the FPGA, and couple it to external controllers in closed-loop HIL and fault tests.

\end{document}